\documentclass{birkjour_t2}
\usepackage{graphicx}
\numberwithin{equation}{section}
\newcommand{\eq}[1]{Eq.~(\ref{#1})}
\newcommand{\eqs}[1]{Eqs.~(\ref{#1})}
\newcommand{\noeq}[1]{(\ref{#1})}
\newcommand{\fig}[1]{Fig.~\ref{#1}}

\newcommand{\be}{\begin{equation}}
\newcommand{\ee}{\end{equation}}
\newcommand{\ba}{\begin{eqnarray}}
\newcommand{\ea}{\end{eqnarray}}
\newcommand{\bs}{\begin{subequations}}
\newcommand{\es}{\end{subequations}}

\newcommand{\tbf}{\textbf}
\newcommand{\mbf}[1]{\mathbf{#1}}
\newcommand{\mcl}[1]{\mathcal{#1}}

\DeclareMathOperator{\Ai}{Ai}
\DeclareMathOperator{\Bi}{Bi}

\begin{document}

\title[Integral representations for products of 
Airy functions with shifted arguments]
{Integral representations for products of two solutions of the
Airy equation with shifted arguments and their applications in physics}

\author[K. V. Bazarov]{Kirill~V.~Bazarov}
\address{Moscow Institute of Physics and Technology,
Dolgoprudny 141700, Russia;\\ NRC “Kurchatov Institute”, Moscow 123182, Russia}
\email{bazarov.kv@phystech.edu}

\author[O. I. Tolstikhin]{Oleg~I.~Tolstikhin}
\address{Moscow Institute of Physics and Technology,
Dolgoprudny 141700, Russia}
\email{tolstikhin.oi@mipt.ru}


\begin{abstract}
Integral representations for a
complete set of linearly independent products of two solutions of 
the Airy equation whose arguments differ by $z_0$ are obtained
using the Laplace contour integral method. 
This generalizes similar integral representations for the case $z_0=0$
obtained by Reid.
The relation to other previous results is discussed. 
The results are used to obtain 
the outgoing-wave Green's function for an electron in a static
electric field in a closed analytic form.
\end{abstract}
\maketitle

\section{Introduction}

Integral representations for various special functions are useful for
applications, particularly as a means to analytically calculate certain
integrals. Although the issue is thoroughly covered in classical
treatises \cite{BE,AS}, new results in this direction continue to
appear. Let $v_1(z)$ and $v_2(z)$ be any solutions of
the Airy equation
\be \label{eq-v}
\left[\frac{d^2}{dz^2}-z\right]v(z)=0.
\ee
Then their product
\be \label{w0}
w(z)=v_1(z)v_2(z)
\ee
satisfies \cite{AS}
\be \label{eq-w0}
\left[\frac{d^3}{dz^3}-4z\frac{d}{dz}
-2\right]w(z)=0.
\ee
This equation can be solved using the Laplace contour integral method \cite{G},
which enables one to obtain integral representations for products of Airy functions.
This program was realized by Reid \cite{reid1995}. It is sufficient to consider
three linearly independent solutions of \eq{eq-w0}, products of any two solutions
of \eq{eq-v} can be expressed in terms of them. Reid obtained integral representations
for $\Ai^2(z)$, $\Ai(z)\Bi(z)$, and $\Bi^2(z)$.
A little later, a related integral representation for the product
$\Ai(u)\Ai(v)$ of two Airy functions
of arbitrary arguments was presented \cite{vallee1997}.
More recently, being motivated
by a physical application and unaware of the results of Reid,
but following the same logic, one of us with colleagues obtained
integral representations for another triple
$\Ai(z)\Ai(e^{\pm 2i\pi/3}z)$ and $\Ai(e^{-2i\pi/3}z)\Ai(e^{+2i\pi/3}z)$
of linearly independent solutions of \eq{eq-w0} \cite{zrp}.
Most recently, working on some exactly solvable models in
the theory of tunneling ionization \cite{gamas},
we have encountered the need to have integral representations for
products of two solutions of the
Airy equation with shifted arguments. In this paper we extend the
approach of Refs.~\cite{reid1995,zrp} to such products and
show that the integral representation obtained in Ref.~\cite{vallee1997}
is a special case of the present results.
We also discuss an application of the present results in physics.

\section{Basic equations}

Let $v_1(z)$ and $v_2(z)$ be any solutions of \eq{eq-v}, as above.
Consider the function
\be \label{w}
w(z;z_0)=v_1(z+z_0)v_2(z),
\ee
where the argument of the first factor is shifted by $z_0$ compared to that in \eq{w0}.
We treat $w(z;z_0)$ as a function of $z$ which depends on $z_0$ as a parameter.
This function satisfies
\be \label{eq-w}
\left[\frac{d^4}{dz^4}-(4z+2z_0)\frac{d^2}{dz^2}
-6\frac{d}{dz}
+z_0^2\right]w(z;z_0)=0.
\ee
We are interested in three particular solutions $\Ai(z)$ and $\Ai(e^{\pm 2i\pi/3}z)$
of \eq{eq-v} which most frequently appear in applications. Using them, one
obtains nine products \noeq{w}. We choose the following
products as four linearly independent solutions of \eq{eq-w}  
\be \label{Upm}
U_{\pm}(z;z_0)=\Ai(e^{\pm 2i\pi/3}(z+z_0))\Ai(e^{\pm 2i\pi/3}z)
\ee
and
\be \label{Wpm}
W_{\pm}(z;z_0)=\Ai(z+z_0)\Ai(e^{\pm 2i\pi/3}z).
\ee
The general solution of \eq{eq-w} is a linear combination of these 
four functions. In particular, the other five products of interest are given by
\bs \label{AA5}
\begin{align}
\Ai(z+z_0)\Ai(z) & = e^{-i\pi/3}W_{+}(z;z_0)+e^{i\pi/3}W_{-}(z;z_0),  \label{2.5a}\\
\Ai(e^{\pm 2i\pi/3}(z+z_0))\Ai(z) & = U_\mp(z;z_0)+
e^{\mp i \pi/3}\left[U_\pm(z;z_0)-W_\mp(z;z_0)\right], \label{2.5b}\\
\Ai(e^{\pm 2i\pi/3}(z+z_0))\Ai(e^{\mp 2i\pi/3}z) & = 
e^{\mp i \pi/3}U_{\mp}(z;z_0)+e^{\pm i \pi/3}W_{\mp}(z;z_0).
\end{align}
\es
Note that all these products are entire functions of $z$ and $z_0$.

Similarly to \eq{eq-w0}, \eq{eq-w} can be solved using the Laplace
contour integral method. We thus obtain particular solutions in the form
\be \label{Ic}
I_{\mcl{C}}(z;z_0)=\int_{\mcl{C}}
\exp\left[i(z+z_0/2)k -\frac{iz_0^2}{4k}
+\frac{ik^3}{12}\right]\frac{dk}{k^{1/2}}.
\ee
The branch of $k^{1/2}$ is determined by the condition that $\arg k=0$ for real 
positive $k$ and a branch cut specified below. 
The integration contour $\mcl{C}$ should be chosen such that the integrand 
multiplied by $k^2$ turns to zero at its ends \cite{G}.
The different solutions of \eq{eq-w} correspond to different contours.
Our goal is to express the solutions \noeq{Upm} and \noeq{Wpm} in terms of such
contour integrals, then integral representations for the other products of interest
will follow from \eqs{AA5}.

The representation of a particular solution by a contour integral can be 
proved by comparing their asymptotics at $|z| \to \infty$. 
In doing so, one should take into account the following.
The general solution of
\eq{eq-w} behaves in the asymptotic region as
\begin{gather}
w(z;z_0) \sim  
\frac{1}{z^{1/2}}\left[c_1\exp\left(\frac{4}{3}z^{3/2}+z_0z^{1/2}\right)+
c_2\exp\left(-\frac{4}{3}z^{3/2}-z_0z^{1/2}\right)
\right. \nonumber \\ 
 \left. +c_3\exp(z_0z^{1/2})+c_4\exp(-z_0z^{1/2})\right]. \label{as}
\end{gather}
This shows that the leading term in the 
asymptotics of a particular solution at a given $\arg z$
does not generally define the solution uniquely. Indeed, consider, for example, 
the ray $\arg z=0$. The most rapidly growing solution that behaves along this 
ray as the first term 
in \eq{as} admits an admixture of subdominant solutions represented by the 
other three terms, and hence is not defined uniquely. However, for any $\arg z$,
the most rapidly decaying solution is uniquely defined by its asymptotics.
For $\arg z=0$ this is the solution that behaves as the second term 
in \eq{as}.

\begin{figure}
\includegraphics[width=90mm]{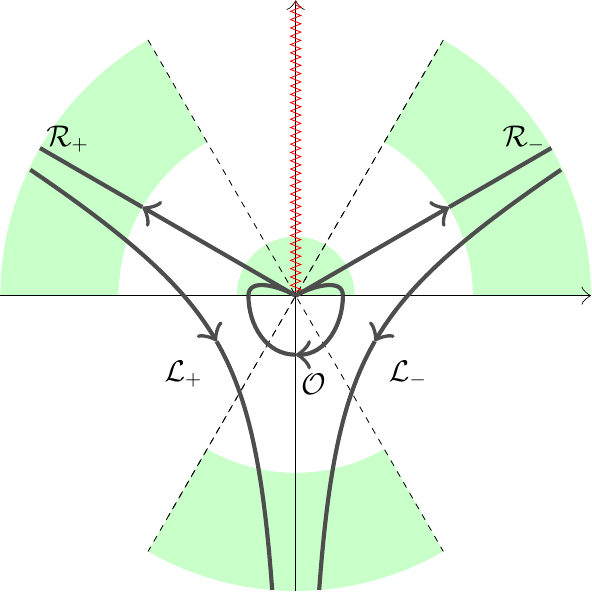}
\caption{\label{f1} The wavy line denotes a branch cut in the complex $k$ plane. 
The solid lines indicate integration contours. 
The colored areas show the internal and asymptotic valleys. 
The branch cut, the contours $\mcl{R}_{\pm}$ and $\mcl{O}$, 
and the internal valley correspond to the case $\arg z_0=0$.}
\end{figure}

\section{Functions $U_{\pm}(z;z_0)$}

Let us make a branch cut at $\arg k=\pi/2$, as shown in \fig{f1}.
The integrand in \eq{Ic} multiplied by $k^2$
decays as $|k| \to \infty$ along three asymptotic valleys 
lying in sectors $-4\pi/3<\arg k<-\pi$, $-2\pi/3<\arg k<-\pi/3$, and
$0<\arg k<\pi/3$. Let contours $\mcl{L}_{\pm}$ go from one of the valleys 
to another, as shown in \fig{f1}. Then
\be \label{Upm-L}
U_{\pm}(z;z_0)=
\frac{e^{i\pi/4\mp i\pi/3}}{4\pi^{3/2}}\,I_{\mcl{L}_{\pm}}(z;z_0).
\ee
To prove this equality, we compare asymptotics of its left- and right-hand
sides at $|z| \to \infty$. 
The asymptotics of $U_{\pm}(z;z_0)$ can be obtained from \eq{Upm}.
These functions present the most rapidly decaying solutions of \eq{eq-w}
along rays $\arg z=\mp 2\pi/3$, respectively, and hence
are uniquely defined by their asymptotics along these rays.
The asymptotics of $I_{\mcl{L}_{\pm}}(z;z_0)$ along these rays
can be found using the saddle-point method \cite{Olver}.
The only saddle point contributing to the integral is
$k=2e^{-i\pi/2 \mp i\pi/3}|z|^{1/2}+O(|z|^{-1/2})$. 
The evaluation of the integral confirms \eq{Upm-L}.
These integral representations hold for any complex $z$ and $z_0$.

\section{Functions $W_{\pm}(z;z_0)$}

The integral representations for $W_{\pm}(z;z_0)$ depend on $\arg z_0$.
We consider the whole interval $-\pi \leq \arg z_0 \leq \pi$ in 
three steps.

\subsection{Ray $\arg z_0=0$}

We begin with the ray $\arg z_0=0$. 
The integrand in \eq{Ic} multiplied by $k^2$ 
decays as $|k| \to 0$ within the internal valley 
lying in sectors $-3\pi/2<\arg k<-\pi$ and 
$0<\arg k<\pi/2$ adjacent to the branch cut.
Let contours $\mcl{R}_{+}$ and $\mcl{R}_{-}$ start at $k=0$
within these sectors and go to 
the corresponding asymptotic valley, as shown in \fig{f1}.
Then
\be \label{Wpm-R}
W_{\pm}(z;z_0)=
\frac{e^{i\pi/4\pm i\pi/3}}{4\pi^{3/2}}I_{\mcl{R}_\pm}(z;z_0).
\ee
To prove this equality for $W_{+}(z;z_0)$ and $W_{-}(z;z_0)$, we compare 
the asymptotics of its left- and right-hand sides as $|z| \to \infty$ in
sectors $-2\pi/3 < \arg z <0$ and $0 < \arg z <2\pi/3$, respectively.
The functions $W_{\pm}(z;z_0)$ behave within these sector as the 
the fourth term in \eq{as}. We first compare the asymptotics 
along rays $\arg z= \mp 0$.
The only saddle point contributing to $I_{\mcl{R}_\pm}(z;z_0)$
is $k=(1/2)e^{-i\pi/2 \mp i\pi \pm i0}z_0|z|^{-1/2}+O(|z|^{-3/2})$.
By evaluating the integral we confirm \eq{Wpm-R}. 
Note, however, that the asymptotics of 
$W_{\pm}(z;z_0)$ along the rays $\arg z= \mp 0$ do not define these functions
uniquely, because there exists a solution $\Ai(z+z_0)\Ai(z)$ 
of \eq{eq-w} that
decays along these rays faster than $W_{\pm}(z;z_0)$. 
This solution could be added to the right-hand side of \eq{Wpm-R}
without changing its asymptotics at $\arg z= \mp 0$.
To resolve this uncertainty, we additionally compare the asymptotics
along rays $\arg z= \mp 2\pi/3 \pm 0$.
The function $\Ai(z+z_0)\Ai(z)$ represents the most rapidly growing solution
along these rays, so adding it to the right-hand side of \eq{Wpm-R}
would violate the equality. This proves \eq{Wpm-R}. 
These integral representations hold for any $z$ under the condition $\arg z_0=0$
assumed in this subsection.

Let $\mcl{O}$ denote a contour which starts and ends at $k=0$ within the internal valley
on the different sides of the branch cut,
as shown in \fig{f1}. It can be seen that
\be \label{I_O}
I_{\mcl{O}}(z;z_0)=I_{\mcl{R}_-}(z;z_0)+I_{\mcl{L}_-}(z;z_0)
-I_{\mcl{L}_+}(z;z_0)-I_{\mcl{R}_+}(z;z_0).
\ee
For $z_0=0$ this contour can be contracted to the point $k=0$, so
the integral $I_{\mcl{O}}(z;z_0)$ turns to zero.
Note that some linear combinations of the solutions \noeq{Upm}-\noeq{AA5}
also turn to zero at $z_0=0$. Integral representations for such linear 
combinations can be conveniently expressed in terms of the contour $\mcl{O}$.
For example, using \eqs{Wpm}, \noeq{2.5b}, \noeq{Upm-L}, \noeq{Wpm-R},
and \noeq{I_O} we obtain
\be \label{A-A}
\Ai(e^{\pm 2i\pi/3}(z+z_0))\Ai(z)-\Ai(z+z_0)\Ai(e^{\pm 2i\pi/3}z)=
\pm \frac{e^{i\pi/4\pm i\pi/3}}{4\pi^{3/2}} I_{\mcl{O}}(z;z_0).
\ee

\subsection{\label{42} Extension to $|\arg z_0|<\pi/2$}

\begin{figure}
\includegraphics[width=65mm]{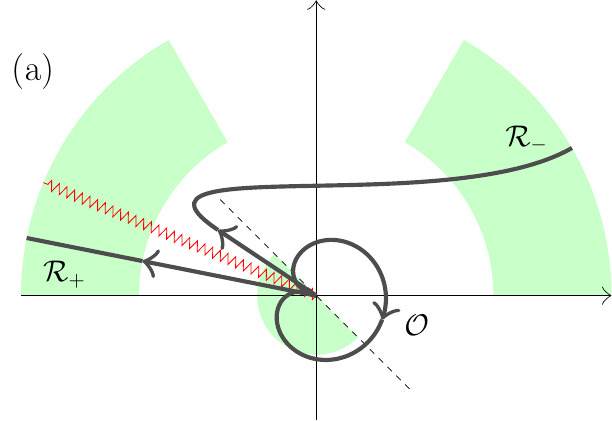}
\hspace{3em}
\includegraphics[width=65mm]{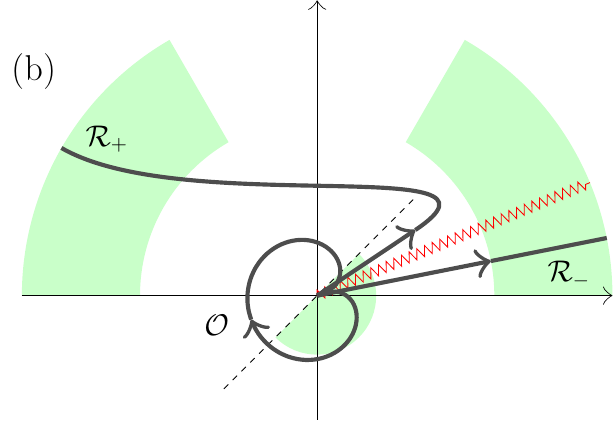}
\caption{\label{f2} The extension of the definition of the
branch cut and the integration contours $\mcl{R}_{\pm}$ and $\mcl{O}$ to:
(a) $0 \leq \arg z_0 < \pi/2$  or  $-\pi \leq \arg z_0 < -\pi/2$,
(b) $-\pi/2 < \arg z_0 \leq 0$  or  $\pi/2 < \arg z_0 \leq \pi$}
\end{figure}

The construction of the previous subsection 
under an appropriate generalization can be extended to the
sector $|\arg z_0|<\pi/2$. 
For a nonzero value of $\arg z_0$, the 
internal valley is rotated with respect to that in \fig{f1} by an angle
$2\arg z_0$, as shown in \fig{f2}. To account for this rotation,
we make the branch cut and the
integration contours dependent on $\arg z_0$ and such that for $\arg z_0=0$
they coincide with those shown in \fig{f1}.
The branch cut is made at $\arg k=\pi/2+\arg z_0$, so that it lies between
the straight-line boundary of the internal valley and the real axis.
The contours $\mcl{R}_{\pm}$ and $\mcl{O}$ in \fig{f2} are obtained by a
continuous
deformation of the corresponding contours in \fig{f1} which
preserves their relative positions with respect to the internal and
asymptotic valleys and the branch cut. In particular, the 
contours $\mcl{R}_{\pm}$ start at $k=0$ within the internal valley 
on the different sides of the branch cut and go to the same asymptotic 
valleys as in \fig{f1}.
Such a deformation of the integration contours implements the
analytic continuation of the right-hand side of \eq{Wpm-R} in $z_0$.
As is clear from \fig{f2}, it is possible only for $|\arg z_0|<\pi/2$.
Thus, with the branch cut and the integration contours dependent on 
$\arg z_0$, as described above, \eq{Wpm-R} remains valid in the sector
$|\arg z_0|<\pi/2$. The same holds for \eqs{I_O} and \noeq{A-A}.

\subsection{Extension to $\pi/2<|\arg z_0| \leq \pi$}

The way how \eq{Wpm-R} was
analytically continued in $z_0$ discussed in the previous
subsection cannot be extended beyond $|\arg z_0|=\pi/2$.
A further extension of the integral representations for
$W_{\pm}(z;z_0)$ to sectors $-\pi \leq \arg z_0 < -\pi/2$
and $\pi/2<\arg z_0 \leq \pi$
can be achieved using the identity
\be
\label{z0mz0}
W_{\pm}(z;z_0)=U_\mp(z;z_0)+
e^{\mp i \pi/3}\left[U_\pm(z;z_0)-W_\mp(z+z_0;-z_0)\right].
\ee
This identity follows from \eqs{Wpm} and \noeq{2.5b}, taking into account that 
$U_\pm(z+z_0;-z_0)=U_\pm(z;z_0)$.
Let us extend the definition of the branch cut and contours $\mcl{R}_{\pm}$ and $\mcl{O}$
as functions of $\arg z_0$
as follows: for a given $z_0$ in sectors $\pi/2 < |\arg z_0| \leq \pi$ they coincide
with those for $-z_0$ 
%in the sector $0 \leq |\arg(- z_0)| < \pi/2$
defined in the previous subsection.
The thus defined branch cut and contours $\mcl{R}_{\pm}$ and $\mcl{O}$
for $|\arg z_0|=\pi$ coincide with those for $\arg z_0=0$ shown in
\fig{f1}. For smaller values of $|\arg z_0|$ in the present interval
they are obtained by a 
continuous deformation, as shown in \fig{f2}.
Such an extension preserves \eq{I_O}. Using this equation, we 
present \eq{z0mz0} in the form
\be \label{Wpm-RO}
W_{\pm}(z;z_0)=
\frac{e^{i\pi/4\pm i\pi/3}}{4\pi^{3/2}}\left[I_{\mcl{R}_\pm}(z;z_0)\pm 
I_{\mcl{O}} (z;z_0)\right].
\ee
Using this,
for the function on the left-hand side of \eq{A-A} we obtain
\be \label{A-A2}
\Ai(e^{\pm 2i\pi/3}(z+z_0))\Ai(z)-\Ai(z+z_0)\Ai(e^{\pm 2i\pi/3}z)=
\mp\frac{e^{i\pi/4\pm i\pi/3}}{4\pi^{3/2}}I_{\mcl{O}} (z;z_0).
\ee
These integral representations hold for any $z$ in sectors 
$\pi/2<|\arg z_0| \leq \pi$.

\subsection{Discussion}

The functions $W_{\pm}(z;z_0)$ are analytic in $z_0$.
Meanwhile, 
%as is clear from the above discussion, 
the branch cut and 
the integration contours $\mcl{R}_{\pm}$ 
defining their integral representations in 
\eqs{Wpm-R} and \noeq{Wpm-RO}
vary continuously
within intervals $|\arg z_0|<\pi/2$ and $\pi/2<|\arg z_0| \leq \pi$,
but undergo an abrupt change corresponding to the jump from
one panel in \fig{f2} to another
at $|\arg z_0| = \pi/2$.
To compensate this change, an additional integral over $\mcl{O}$ appears in
\eq{Wpm-RO}. As follows from \eqs{A-A} and \noeq{A-A2},
this additional term vanishes at $z_0=0$.
For $z_0 \ne 0$, it accounts for the difference between 
$W_{\pm}(z;z_0)$ and $W_{\pm}(z+z_0;-z_0)$.
Finally, we mention that the integral representations for $W_{\pm}(z;z_0)$
at the rays $\arg z_0=\pm \pi/2$ can be obtained by continuity in $z_0$
in the corresponding limit
from either \eq{Wpm-R} or \eq{Wpm-RO}.

\section{Real $z$ and $z_0$}

Consider real $z$ and $z_0$ denoted by $x$ and $x_0$, respectively.
In this case contours $\mcl{R}_{\pm}$ can be placed on the real axis
and some of the formulas discussed above can be simplified.
For $x_0 \geq 0$, we obtain from \eq{Wpm-R}
\be \label{W-x}
W_{\pm}(x,x_0)=\frac{e^{\pm i\pi/12}}{4\pi^{3/2}}\int_0^\infty 
\exp\left[\mp i k(x+x_0/2)\pm i\frac{x_0^2}{4k}\mp i\frac{k^3}{12}\right] 
\frac{dk}{k^{1/2}}.
\ee
If $x_0=0$, this formula reduces to the integral representation 
for $\Ai(z)\Ai(e^{\pm 2i\pi/3}z)$ obtained in Ref.~\cite{zrp}.
Note that such a simple formula cannot be written for $x_0 < 0$,
because of the term involving the contour $\mcl{O}$ in \eq{Wpm-RO}.
Using \eq{W-x}, we obtain from \eq{2.5a}
\be \label{AA-x}
\Ai(x+x_0)\Ai(x)  = \frac{1}{2\pi^{3/2}}\int_0^\infty 
\cos\left[k(x+x_0/2)-\frac{x_0^2}{4k}+\frac{k^3}{12}+\frac{\pi}{4}\right] 
\frac{dk}{k^{1/2}}.
\ee
Remarkably, this formula holds for any real $x$ and $x_0$, including the
case $x_0<0$ where \eq{W-x} does not hold, because the integral over 
$\mcl{O}$ in \eq{Wpm-RO} is canceled upon substituting into \eq{2.5a}.
Equation \noeq{AA-x} with $x=v$ and $x_0=u-v$
reproduces the integral representation for $\Ai(u)\Ai(v)$
obtained in Ref.~\cite{vallee1997}. We conclude that this integral 
representation holds only for real $u$ and $v$, which was not 
indicated in Ref.~\cite{vallee1997}, 
and its validity 
within the present analysis relies on a nontrivial cancellation 
of the integral over $\mcl{O}$ in \eq{Wpm-RO}.

\section{An application: 
Green's function for an electron in a static electric field}

Integrals of the type \noeq{Ic} arise in many physical applications. 
In particular, such an integral defines 
the outgoing-wave Green's function for an electron in a static
electric field, which plays an important role, e.g., 
in the theory of tunneling ionization \cite{gamas,frp}.
This function is the outgoing-wave solution of
the stationary Schr\"{o}dinger equation
(here we use atomic units $\hbar=m=e=1$)
\be
\left[-
\frac{1}{2}\,\Delta+\mbf{F}\mbf{r}-E\right]
G(\mbf{r},\mbf{r}')=\delta(\mbf{r}-\mbf{r}'),
\ee
where $E$ is the electron energy and $\mbf{F}$
is the electric field.
The solution can be obtained by Fourier transforming
the corresponding retarded Green's function, which can be
found analytically, and is commonly presented in the form \cite{frp}
\be \label{G-int}
G(\mbf{r},\mbf{r}')=
\frac{e^{-i\pi/4}}{(2\pi)^{3/2}}
\int_0^{\infty}
\exp\left[iEt+\frac{i(\mbf{r}-\mbf{r}')^2}{2t}-
\frac{i}{2}\,\mbf{F}(\mbf{r}+\mbf{r}')t
-\frac{i}{24}\,\mbf{F}^2t^3\right]\frac{dt}{t^{3/2}}.
\ee
This integral can be calculated using \eq{W-x}.
Let us introduce new variables
\be
\xi=\frac{\mbf{F}(\mbf{r}+\mbf{r}')-2E}{(2F)^{2/3}},\quad
\eta=\frac{F^{1/3}}{2^{2/3}}|\mbf{r}-\mbf{r}'|,
\ee
where $F=|\mbf{F}|$. Note that $\eta$ is real and positive, 
which is important for the applicability of \eq{W-x}. 
Comparing \eq{G-int} with \eq{W-x}, we obtain
\be \label{G}
G(\mbf{r},\mbf{r}')=
\frac{-e^{i\pi/6}}{|\mbf{r}-\mbf{r}'|}
\frac{d}{d\eta}\Ai(\xi+\eta)\Ai(e^{2i\pi/3}(\xi-\eta)).
\ee
As far as we know, such a closed analytic form for
this Green's function has never been presented in
the literature. Equation \noeq{G} enables one to extend the 
theory of tunneling ionization 
\cite{gamas} to strong fields beyond the weak-field
limit, which is very important for applications in attosecond
physics \cite{KI}.
In the limit $F \to 0$ \eq{G} reduces to
\be
G(\mbf{r},\mbf{r}')=
\frac{e^{ik|\mbf{r}-\mbf{r}'|}}{2\pi|\mbf{r}-\mbf{r}'|},
\ee
where $k=\sqrt{2E}$. This is the well-known Green's function
for a free electron.

\section{Conclusion}

We have obtained integral representations for a complete set 
of four linearly independent
solutions \noeq{Upm} and \noeq{Wpm} of \eq{eq-w} given by
\eqs{Upm-L}, \noeq{Wpm-R}, and \noeq{Wpm-RO}.
Any product \noeq{w} of two solutions of 
the Airy equation \noeq{eq-v} whose arguments are shifted by $z_0$ can be expressed
in terms of these four functions, as in \eqs{AA5}, and hence
our results enable one to obtain an integral representation
for any such product. 
This generalizes the work of Reid \cite{reid1995}
who realized a similar program for the case $z_0=0$.
Using these results, we have obtained 
the outgoing-wave Green's function for an electron in a static
electric field in a closed analytic form, \eq{G}, which is important
for many applications in physics.

\subsection*{Acknowledgment}
This work supported by the Ministry of Science and Higher Education of
the Russian Federation (No.~FSMG-2021-0005).

\end{document}